\documentstyle[aps,epsfig,12pt]{revtex}

\begin{document}
ADP-05-12/T622

JLAB-THY-05-312
\begin{center}
\vspace*{1cm} {\Large {\bf Neutron stars and strange stars in the chiral
{\em\bf SU(3)} quark mean field model}}\\[0pt]
\vspace*{1cm} P. Wang$^{ab}$, S. Lawley$^b$, D. B. Leinweber$^a$, A. W. Thomas$^b$
and A. G. Williams$^a$ \\[0pt]
\vspace*{0.2cm} {\it $^a$Special Research Center for the Subatomic Structure of
Matter (CSSM) and Department of Physics, University of Adelaide 5005, Australia } \\[0pt]
\vspace*{0.2cm} {\it $^b$Jefferson Laboratory, 12000 Jefferson Ave., Newport News, VA 23606 USA}\\[0pt]
\end{center}

\begin{abstract}

We investigate the equations of state for pure neutron matter and strange
hadronic matter in $\beta$-equilibrium, including $\Lambda$,
$\Sigma$ and $\Xi$ hyperons. The
masses and radii of pure neutron stars and strange hadronic stars
are obtained. For a pure neutron star, the maximum mass is
about $1.8 M_{\mathrm{sun}}$, while for a strange hadronic star, the
maximum mass is around $1.45 M_{\mathrm{sun}}$. The typical radii of pure
neutron stars and strange hadronic stars are about 11.0-12.3 km
and 10.7-11.7 km, respectively.

\end{abstract}

\bigskip

\leftline{PACS number(s): 21.65.+f; 26.60.+c; 21.80+a; 12.39.-x}
\bigskip
\leftline{{\bf Keywords: Hadronic Matter, Neutron Star, Strange Star,
Chiral Symmetry,}}

{\bf ~~~~~~~~~~ Quark Mean Field}

\section{Introduction}

Hadronic matter under extreme conditions has attracted a lot of
interest in recent years. On the one hand, many theoretical and
experimental efforts
have been devoted to the discussion of heavy ion collisions where the
temperature is high. On the other hand, the physics of neutron
stars has become a hot topic which connects astrophysics with high
density nuclear physics. In 1934, Baade and Zwicky \cite{Baade} suggested that
neutron stars could be formed in supernovae.
The first theoretical calculation
of a neutron star was performed by Oppenheimer and Volkoff \cite{Oppenheimer},
and independently by Tolman \cite{Tolman}. Observing a range of masses and
radii of neutron stars will
reveal the equations of state (EOS) of dense hadronic matter.
Determination of the EOS of neutron stars has been an important
goal for more than two decades. Six
double neutron-star binaries are known so far, and all of them
have masses in the surprisingly narrow range $1.36 \pm 0.08M_{\mathrm{sun}}$
\cite{Thorsett,Heiselberg}. A number of early theoretical investigations
on neutron stars were based on the non-relativistic Skyrme framework
\cite{Bonche}.
Since the Walecka model \cite{Serot} was proposed and applied to study
the properties of nuclear matter, the relativistic mean field approach has
been widely used in the determination of the masses and radii of neutron stars.
These models lead to different predictions for neutron
star masses and radii \cite{Ban,Lawley}. For a recent review see
Ref.\ \cite{Weber}. Though models with maximum neutron star masses
considerably smaller than $1.4M_{\mathrm{sun}}$ are simply ruled out,
the constraint on EOS of nuclear matter (for example, the density
dependence of pressure of hadronic system) has certainly not been
established from the existing observations.

In the process of neutron star formation,
$\beta$-equilibrium can be achieved. As a consequence, hyperons will
exist in neutron stars, especially in stars with high baryon
density. These hyperons will affect the EOS of hadronic matter. As
a result, the mass-radius relationship of strange hadronic stars will
be quite different from that of pure neutron stars. The simplest way to
discuss the effects of hyperons is to study strange hadronic stars
including only $\Lambda$ hyperons. This is due to the fact that $\Lambda$ is the
lightest hyperon and the $\Lambda$-N interaction is known better
than other hyperon-nucleon interactions. However, one must also consider
hyperons with negative charge in neutron stars because the
negatively charged hyperons can substitute for electrons. There
have been many discussions of strange hadronic stars including $\Lambda$
hyperons, $\Lambda$ and $\Sigma^-$ or even the whole baryon octet
\cite{Nishizaki} -\cite{Maruyama}.

At high baryon density, the overlap effects of baryons are very
important and the quark degrees of freedom within baryons should be considered.
There are some phenomenological models based on the quark degrees of freedom,
such as the quark meson coupling model \cite{Guichon},
the cloudy bag model \cite{Thomas}, the quark mean field model \cite{Toki}
and the NJL model \cite{Bentz}.
Several years ago, a chiral $SU(3)$ quark mean field model was
proposed \cite{Wang3,Wang4}.
In this model, quarks are confined within baryons by an effective
potential. The quark-meson interaction and meson self-interaction
are based on $SU(3)$ chiral symmetry. Through the mechanism of
spontaneous symmetry breaking the resulting constituent quarks
and mesons (except for the pseudoscalars) obtain masses. The
introduction of an explicit symmetry breaking term in the meson
self-interaction generates the masses of the pseudoscalar mesons
which satisfy the partially conserved axial-vector current
(PCAC) relations. The explicit symmetry
breaking term in the quark-meson interaction gives reasonable
hyperon potentials in hadronic matter. This chiral $SU(3)$ quark
mean field model has been applied to investigate nuclear matter \cite{Wang2},
strange hadronic matter \cite{Wang3}, finite nuclei, hypernuclei \cite{Wang4},
and quark matter \cite{Wang5}. Recently, we improved the chiral
$SU(3)$ quark mean field model by using the linear definition of
effective baryon mass \cite{Wangcssm1}. This new treatment is
applied to study the liquid-gas phase transition of asymmetric
nuclear system and strange hadronic matter \cite{Wangcssm2,Wangcssm3}.
By and large the results are in reasonable agreement with existing
experimental data.

In this paper, we will study the neutron star and strange star in
the chiral $SU(3)$ quark mean field model. The paper is organized in
the following way. In section II, we briefly introduce the model.
In section III, we apply this model to investigate the neutron
star and strange hadronic star. The numerical results are discussed in
section IV. We summarize the main results in section V.

\section{The model}

Our considerations are based on the chiral $SU(3)$ quark mean field model
(for details see Refs.~\cite{Wang3,Wang4}), which contains
quarks and mesons as the basic degrees of freedom.
In the chiral limit, the quark field $\Psi$ can be split into left and
right-handed parts $\Psi_{L}$ and $\Psi_{R}$: $\Psi=\Psi_{L}+\Psi_{R}$.
Under $SU(3)$$_{L}\times$ $SU(3)$$_{R}$ they transform as
\begin{equation}
\Psi_{L} \rightarrow \Psi_{L}^{\prime }\,=\,L\,\Psi_{L},~~~~~
\Psi_{R} \rightarrow \Psi_{R}^{\prime }\,=\,R\,\Psi_{R}\,.
\end{equation}
The spin-0 mesons are written in the compact form
\begin{equation}
{M \atop M^\dag}=\Sigma \pm i\Pi =\frac{1}{\sqrt{2}}\sum_{a=0}^{8}
\left( s^{a}\pm i p ^{a}\right) \lambda ^{a},
\end{equation}
where $s^{a}$ and $p ^{a}$ are the nonets of scalar and pseudoscalar
mesons, respectively, $\lambda ^{a}(a=1,...,8)$ are the Gell-Mann
matrices, and $\lambda ^{0}=\sqrt{\frac{2}{3}}\,I$. The alternatives
indicated by the
plus and minus signs correspond to $M$ and $M^\dag$, respectively.
Under chiral $SU(3)$ transformations, $M$ and $M^\dag$ transform as
$M\rightarrow M^{\prime }=LMR^\dag$ and $M^\dag\rightarrow
M^{\dag^{\prime }}=RM^{\dag}L^{\dag}$. The spin-1
mesons are arranged in a similar way as
\begin{equation}
{l_{\mu } \atop r_{\mu }}=\frac{1}{2}\left( V_{\mu }\pm A_{\mu }\right)
= \frac{1}{2\sqrt{2}}\sum_{a=0}^{8}\left( v_{\mu }^{a}\pm a_{\mu }^{a}
\right) \lambda^{a}
\end{equation}
with the transformation properties:
$l_{\mu}\rightarrow l_{\mu }^{\prime }=Ll_{\mu }L^{\dag}$,
$r_{\mu}\rightarrow r_{\mu }^{\prime }=Rr_{\mu }R^{\dag}$.
The matrices $\Sigma$, $\Pi$, $V_{\mu }$ and $A_{\mu }$ can be
written in a form where the physical states are explicit.
For the scalar and vector nonets, we have the expressions
\begin{eqnarray}
\Sigma = \frac1{\sqrt{2}}\sum_{a=0}^8 s^a \, \lambda^a=\left(
\begin{array}{lcr}
\frac1{\sqrt{2}}\left(\sigma+a_0^0\right) & a_0^+ & K^{*+} \\
a_0^- & \frac1{\sqrt{2}}\left(\sigma-a_0^0\right) & K^{*0} \\
K^{*-} & \bar{K}^{*0} & \zeta
\end{array}
\right),
\end{eqnarray}
\begin{eqnarray}
V_\mu = \frac1{\sqrt{2}}\sum_{a=0}^8 v_\mu^a \, \lambda^a=\left(
\begin{array}{lcr}
\frac1{\sqrt{2}}\left(\omega_\mu+\rho_\mu^0\right)
& \rho_\mu^+ & K_\mu^{*+}\\
\rho_\mu^- & \frac1{\sqrt{2}}\left(\omega_\mu-\rho_\mu^0\right)
& K_\mu^{*0}\\
K_\mu^{*-} & \bar{K}_\mu^{*0} & \phi_\mu
\end{array}
\right).
\end{eqnarray}
Pseudoscalar and pseudovector nonet mesons can be written in
a similar fashion.

The total effective Lagrangian is written:
\begin{eqnarray}
{\cal L}_{{\rm eff}} \, = \, {\cal L}_{0} \, + \, {\cal L}_{qM}
\, + \,
{\cal L}_{\Sigma\Sigma} \,+\, {\cal L}_{VV} \,+\, {\cal L}_{\chi SB}\,
+ \, {\cal L}_{\Delta m_s} \, + \, {\cal L}_{h}, + \, {\cal L}_{c},
\end{eqnarray}
where ${\cal L}_{0} =i\bar{\Psi}\gamma^\mu \partial_\mu \Psi$ is the
free part for massless quarks. The quark-meson interaction
${\cal L}_{qM}$ can be written in a chiral $SU(3)$ invariant way as
\begin{eqnarray}
{\cal L}_{qM}=g_s\left(\bar{\Psi}_LM\Psi_R+\bar{\Psi}_RM^\dag\Psi_L\right)
-g_v\left(\bar{\Psi}_L\gamma^\mu l_\mu\Psi_L+\bar{\Psi}_R\gamma^\mu
r_\mu\Psi_R\right)~~~~~~~~~~~~~~~~~~~~~~~  \nonumber \\
=\frac{g_s}{\sqrt{2}}\bar{\Psi}\left(\sum_{a=0}^8 s_a\lambda_a
+ i \gamma^5 \sum_{a=0}^8 p_a\lambda_a
\right)\Psi -\frac{g_v}{2\sqrt{2}}
\bar{\Psi}\left( \gamma^\mu \sum_{a=0}^8
 v_\mu^a\lambda_a
- \gamma^\mu\gamma^5 \sum_{a=0}^8
a_\mu^a\lambda_a\right)\Psi.
\end{eqnarray}
From the quark-meson interaction, the coupling constants between
scalar mesons, vector mesons and quarks have the following
relations:
\begin{eqnarray}
\frac{g_s}{\sqrt{2}}
&=& g_{a_0}^u = -g_{a_0}^d = g_\sigma^u = g_\sigma^d = \ldots =
\frac{1}{\sqrt{2}}g_\zeta^s, \label{relation}
~~~~~g_{a_0}^s = g_\sigma^s = g_\zeta^u = g_\zeta^d = 0 \, ,\\
\frac{g_v}{2\sqrt{2}}
&=& g_{\rho}^u = -g_{\rho}^d = g_\omega^u = g_\omega^d = \ldots =
\frac{1}{\sqrt{2}}g_\phi^s,
~~~~~g_\omega^s = g_{\rho}^s = g_\phi^u = g_\phi^d = 0 .
\end{eqnarray}

In the mean field approximation, the chiral-invariant scalar meson
${\cal L}_{\Sigma\Sigma}$ and vector meson ${\cal L}_{VV}$
self-interaction terms are written as~\cite{Wang3,Wang4}
\begin{eqnarray}
{\cal L}_{\Sigma\Sigma} &=& -\frac{1}{2} \, k_0\chi^2
\left(\sigma^2+\zeta^2\right)+k_1 \left(\sigma^2+\zeta^2\right)^2
+k_2\left(\frac{\sigma^4}2 +\zeta^4\right)+k_3\chi\sigma^2\zeta
\nonumber \\ \label{scalar}
&&-k_4\chi^4-\frac14\chi^4 {\rm ln}\frac{\chi^4}{\chi_0^4} +
\frac{\delta}
3\chi^4 {\rm ln}\frac{\sigma^2\zeta}{\sigma_0^2\zeta_0}, \\
{\cal L}_{VV}&=&\frac{1}{2} \, \frac{\chi^2}{\chi_0^2} \left(
m_\omega^2\omega^2+m_\rho^2\rho^2+m_\phi^2\phi^2\right)+g_4
\left(\omega^4+6\omega^2\rho^2+\rho^4+2\phi^4\right), \label{vector}
\end{eqnarray}
where $\delta = 6/33$; $\sigma_0$ and $\zeta_0$ are the
vacuum expectation values of the corresponding mean fields
$\sigma$, $\zeta$ which are expressed as
\begin{equation}
\sigma_0=-F_\pi, ~~~~ \zeta_0=\frac1{\sqrt{2}}(F_\pi-2F_K).
\end{equation}
The vacuum value $\chi_0$ is about 280 MeV in our numerical
calculation.
The Lagrangian ${\cal L}_{\chi SB}$ generates the
nonvanishing masses of pseudoscalar mesons
\begin{equation}\label{L_SB}
{\cal L}_{\chi SB}=\frac{\chi^2}{\chi_0^2}\left[m_\pi^2F_\pi\sigma +
\left(
\sqrt{2} \, m_K^2F_K-\frac{m_\pi^2}{\sqrt{2}} F_\pi\right)\zeta\right],
\end{equation}
leading to a nonvanishing divergence of the axial currents which in
turn satisfy the partial conserved axial-vector current (PCAC)
relations for $\pi$ and $K$ mesons. Pseudoscalar,
scalar mesons and also the dilaton field $\chi$ obtain mass terms by
spontaneous breaking of chiral symmetry in the Lagrangian of Eq.~(\ref{scalar}).
The masses of $u$, $d$ and $s$ quarks are generated by
the vacuum expectation values of the two scalar mesons $\sigma$ and
$\zeta$. To obtain the correct constituent mass of the strange quark,
an additional mass term has to be added:
\begin{eqnarray}
{\cal L}_{\Delta m_s} = - \Delta m_s \bar q S q
\end{eqnarray}
where $S \, = \, \frac{1}{3} \, \left(I - \lambda_8\sqrt{3}\right) =
{\rm diag}(0,0,1)$ is the strangeness quark matrix. Based on these
mechanisms, the quark constituent masses are finally given by
\begin{eqnarray}
m_u=m_d=-\frac{g_s}{\sqrt{2}}\sigma_0
\hspace*{.5cm} \mbox{and} \hspace*{.5cm}
m_s=-g_s \zeta_0 + \Delta m_s.
\end{eqnarray}
The parameters $g_s = 4.76$ and $\Delta m_s = 29$~MeV are chosen to yield
the constituent quark masses $m_q=313$~MeV and $m_s=490$~MeV.
In order to obtain reasonable hyperon potentials in hadronic matter, we
include an additional coupling between strange quarks and the scalar
mesons $\sigma$ and $\zeta$~\cite{Wang3}. This term is expressed as
\begin{eqnarray}
{\cal L}_h \,=\, [h_1 (\sigma-\sigma_0)+h_2(\zeta-\zeta_0)] \, \bar{s} s \,.
\end{eqnarray}
Therefore, the strange quark scalar-coupling constants are modified
and do not exactly satisfy Eq.~(\ref{relation}).
The hyperon potentials were listed in our previous paper
\cite{Wangcssm1}.
In the quark mean field model, quarks are confined in baryons
by the Lagrangian ${\cal L}_c=-\bar{\Psi} \, \chi_c \, \Psi$ (with $\chi_c$
given in Eq.~(\ref{Dirac}), below).
We note that this confining term is not chiral invariant. Possible
extensions of the model which would restore chiral symmetry in this term
have been discussed in Ref.~\cite{confining}.

The Dirac equation for a quark field $\Psi_{ij}$ under the additional
influence of the meson mean fields is given by
\begin{equation}
\left[-i\vec{\alpha}\cdot\vec{\nabla}+\beta \chi_c(r)+\beta m_i^*\right]
\Psi_{ij}=e_i^*\Psi_{ij}, \label{Dirac}
\end{equation}
where $\vec{\alpha} = \gamma^0 \vec{\gamma}$\,, $\beta = \gamma^0$\,,
the subscripts $i$ and $j$ denote the quark $i$ ($i=u, d, s$)
in a baryon of type $j$ ($j=N, \Lambda, \Sigma, \Xi$)\,;
$\chi_c(r)$ is a confinement potential, i.e. a static potential
providing the confinement of quarks by meson mean-field configurations.
In the numerical calculations, we choose $\chi_{c}(r)= \frac14
k_cr^2$, where $k_{c} = 1$ (GeV fm$^{-2})$, which yields baryon radii
(in the absence of the pion cloud \cite{Hackett}) around 0.6 fm.
The quark mass $m_i^*$ and energy $e_i^*$ are defined as
\begin{equation}
m_i^*=-g_\sigma^i\sigma - g_\zeta^i\zeta+m_{i0}
\end{equation}
and
\begin{equation}
e_i^*=e_i-g_\omega^i\omega-g_\rho^i\rho-g_\phi^i\phi \,,
\end{equation}
where $e_i$ is the energy of the quark under the influence of
the meson mean fields. Here $m_{i0} = 0$ for $i=u,d$ (nonstrange quark)
and $m_{i0} = \Delta m_s$ for $i=s$ (strange quark).
The effective baryon mass can be written as
\begin{eqnarray}
M_j^*=\sum_in_{ij}e_i^*-E_j^0 \label{linear}\,,
\end{eqnarray}
where $n_{ij}$ is the number of quarks with flavor $``i"$ in a baryon
with flavor $j$, with $j = N \, \{p, n\}\,,
\Sigma \, \{\Sigma^\pm, \Sigma^0\}\,, \Xi \,\{\Xi^0, \Xi^-\}\,,
\Lambda\,$  and $E_j^0$ was found to be only very weakly dependent on the
external field strength.
We therefore use Eq.~(\ref{linear}), with $E_j^0$ a
constant, independent of the density, which is adjusted to give a
best fit to the free baryon masses. Compared with the earlier
square root ans\"atz, here we use the linear
definition of effective baryon mass. As we have explained in Ref.
\cite{Wangcssm1} the linear definition of effective baryon mass has
been derived using a symmetric relativistic approach \cite{Guichon2},
while to the best of our
knowledge, no equivalent derivation exists for the square
root case.

\section{hadronic system}

Based on the previously defined quark mean field model
the thermodynamical potential for the study of hadronic
systems is written as
\begin{equation}
\Omega=\sum_{j=N,\Lambda,\Sigma,\Xi}\frac{-2k_BT}{(2\pi)^3}\int_0^\infty
d^3k\left\{ {\rm ln}
\left(1+e^{-(E_j^*(k)-\nu_j)/k_BT}\right)+
{\rm ln} \left(1+e^{-(E_j^*(k)+\nu_j)/k_BT}\right)\right\} -
{\cal L}_M,
\end{equation}
where $E_j^*(k)=\sqrt{M_j^{*2}+k^2}$ and $M_j^*$ is the effective
baryon mass. The quantity
$\nu_j$ is related to the usual chemical potential
$\mu_j$ by $\nu_j=\mu_j-g_\omega^j\omega-g_\rho^j\rho-g_\phi^j\phi$.
The mesonic Lagrangian
\begin{equation}
{\cal L}_M = {\cal L}_{\Sigma\Sigma} + {\cal L}_{VV}
+ {\cal L}_{\chi SB}
\end{equation}
describes the interaction between mesons which
includes the scalar meson self-interaction ${\cal L}_{\Sigma\Sigma}$,
the vector meson self-interaction ${\cal L}_{VV}$ and the explicit
chiral symmetry breaking term ${\cal L}_{\chi SB}$ defined
previously in Eqs.~(\ref{scalar}), (\ref{vector}) and (\ref{L_SB}).
The Lagrangian ${\cal L}_M$ involves scalar ($\sigma$, $\zeta$
and $\chi$) and vector ($\omega$, $\rho$ and $\phi$) mesons.
The interactions
between quarks and scalar mesons result in the effective baryon masses
$M_j^*$.
The interactions between quarks and vector mesons generate the
baryon-vector meson interaction terms.
The energy per volume and the pressure of the system can be
derived as $\varepsilon =\Omega -\frac1T
\frac{\partial\Omega}{\partial T}+\nu _j\rho_j$ and $p=-\Omega $,
where $\rho_j$ is the density of baryon $j$.
At zero temperature, $\Omega$ can be expressed as
\begin{eqnarray}
\Omega &=& -\sum\limits_{j=N,\Lambda,\Sigma,\Xi}\frac 1{24\pi^2}
\left\{\nu_j\left[\nu_j^2-M_j^{*2}\right]^{1/2}
\left[2\nu_j^2-5M_j^{*2}\right] \right .
\nonumber\\
&+& \left . 3M_j^{*4}{\rm ln}\left[\frac{\nu_j+\left(\nu_j^2-
M_j^{*2}\right)^{1/2}}{M_j^*}\right]\right\} - {\cal L}_M,
\end{eqnarray}

The equations for mesons $\phi_{i}$ can be obtained by
the formula $\frac{\partial\Omega}{\partial \phi_i}=0$.
Therefore, the equations for $\sigma$, $\zeta$ and $\chi$ are
\begin{eqnarray}
k_0\chi^2\sigma-4k_1\left(\sigma^2+\zeta^2 \right)\sigma-2k_2\sigma^3
-2k_3\chi\sigma\zeta-\frac{2\delta}{3\sigma}\chi^4+\frac{\chi^2}{\chi_0^2}
m_\pi^2F_\pi \nonumber \\
-\left(\frac{\chi}{\chi_0}\right)^2m_\omega\omega^2\frac{\partial m_\omega}
{\partial\sigma}
-\left(\frac{\chi}{\chi_0}\right)^2m_\rho\rho^2\frac{\partial m_\rho}
{\partial\sigma}+\sum_{j = N\,, \Lambda\,, \Sigma\,, \Xi }
\frac{\partial M_{j}^{\ast }}{\partial \sigma } <\bar{\psi _{j}}\psi_{j}>&=&0,
\end{eqnarray}
\begin{eqnarray}
k_0\chi^2\zeta-4k_1\left(\sigma^2+\zeta^2\right)\zeta-4k_2\zeta^3
-k_3\chi\sigma^2-\frac{\delta}{3\zeta}\chi^4+\frac{\chi^2}{\chi_0^2}
\left(\sqrt{2}m_k^2F_k-\frac1{\sqrt{2}}m_\pi^2F_\pi\right) \nonumber \\
-\left(\frac{\chi}{\chi_0}\right)^2m_\phi\phi^2\frac{\partial m_\phi}
{\partial\zeta}+\sum_{j = \Lambda\,, \Sigma\,, \Xi}
\frac{\partial M_{j}^{\ast}}{\partial\zeta}
 <\bar{\psi_{j}}\psi_{j}>=0,~~~~~~~~~~~~~~~~~~~
\end{eqnarray}
\begin{eqnarray}
k_0\chi\left(\sigma^2+\zeta^2\right)-k_3\sigma^2\zeta+\left(4k_4+1
+4ln\frac{\chi}{\chi_0}-\frac{4\delta}{3}ln\frac{\sigma^2\zeta}{\sigma_0^2
\zeta_0}
\right)\chi^3 \nonumber \\
+\frac{2\chi}{\chi_0^2}\left[m_\pi^2F_\pi\sigma+
\left(\sqrt{2}m_k^2F_k-\frac1{\sqrt{2}}m_\pi^2F_\pi\right)\zeta\right]
-\frac{\chi}{\chi_0^2}(m_\omega^2\omega^2+m_\rho^2\rho^2+m_\phi^2\phi^2)&=&0,
\end{eqnarray}
where $<\bar{\psi}_j \psi_j>$ is expressed as
\begin{eqnarray}
<\bar{\psi}_j \psi_j>&=&\frac{M_j^\ast}{\pi^2} \,
\int_{0}^{k_{F_j}} dk \frac{k^2}{\sqrt{M_j^{\ast 2}+k^2}} \\  \nonumber
&=& \frac{M_j^{\ast 3}}{2 \, \pi^2} \,
\biggl[ \frac{k_{F_j}}{M_j^\ast} \,
\sqrt{1 + \frac{k_{F_j}^2}{M_j^{\ast 2}}}
 -  {\rm ln}\biggl( \frac{k_{F_j}}{M_j^\ast} +
\sqrt{1 + \frac{k_{F_j}^2}{M_j^{\ast 2}}} \biggr) \biggr],
\end{eqnarray}
with $k_{F_j}=\sqrt{\nu_j^2-M_j^{*2}}$.

For the $\beta$-equilibrium, the chemical potentials for the baryons
satisfy the following equations:
\begin{equation}
\mu_\Lambda=\mu_{\Sigma^0}=\mu_{\Xi^0}=\mu_n=\mu_p+\mu_e=\mu_p+\mu_\mu,
\end{equation}
\begin{equation}
\mu_{\Sigma^+}=\mu_p,
\end{equation}
\begin{equation}
\mu_{\Sigma^-}=\mu_{\Xi^-}=\mu_n+\mu_e.
\end{equation}
There are only two independent chemical potentials which are determined by the total
baryon density and neutral charge:
\begin{equation}
\rho_B=\rho_p+\rho_n+\rho_\Lambda+\rho_{\Sigma^+}+\rho_{\Sigma^0}+
\rho_{\Sigma^-}+\rho_{\Xi^0}+\rho_{\Xi^-},
\end{equation}
\begin{equation}
\rho_p+\rho_{\Sigma^+}-\rho_{\Sigma^-}-\rho_{\Xi^-}-\rho_e-\rho_\mu=0.
\end{equation}

In order to get the mass-radius relation, one has to resolve the
Tolman-Oppenheimer-Volkoff (TOV) equation:
\begin{equation}
\frac{dp}{dr}=-\frac{\left[p(r)+\varepsilon(r)\right]\left[M(r)+4\pi r^3p(r)\right]}
{r(r-2M(r))},
\end{equation}
where
\begin{equation}
M(r)=4\pi\int_0^r \varepsilon(r) r^2 dr.
\end{equation}
With the equations of state, the functions, such as $M(r)$, $\rho(r)$ and
$p(r)$, etc. can be obtained.

\section{Numerical results}

The parameters of this model were determined by the meson masses
in vacuum and the saturation properties of nuclear matter which
were listed in the table 1 of Ref. \cite{Wangcssm1}. The improved linear
definition of effective baryon mass is chosen in our numerical calculations.
We first discuss the equations of
state of neutron matter and strange hadronic matter which are
needed for the calculation of neutron stars. For pure neutron
stars, there are only neutrons present. For strange hadronic stars,
with increasing baryon density, other kinds of baryons
will appear. In Fig.~1 we show the fractions of octet baryons versus density
with $\beta$-equilibrium. With the increasing of baryon density, the
neutron fraction decreases slowly from 1. If the density is lower
than about 0.19 fm$^{-3}$, the fraction of electrons is the same as that
of protons which makes the system charge neutral. The muon appears
when the density is in the range 0.19 - 0.98 fm$^{-3}$. The maximum
fractions of muons and electrons appear at $\rho_B\simeq$ 0.4 fm$^{-3}$.
Their fractions decrease with the increasing fractions of hyperons.
When the density is larger than about
0.4 fm$^{-3}$, the $\Sigma^-$ hyperons appear and the fraction of
neutrons decreases faster. After the density is larger than about
0.57 fm$^{-3}$, $\Lambda$ hyperons start to appear.
The fraction of $\Sigma^-$ hyperons decreases with the increasing density
after $\Xi^-$ hyperons appear where the density is about 0.84 fm$^{-3}$.

\begin{center}
\begin{figure}[hbt]
\includegraphics[scale=0.66]{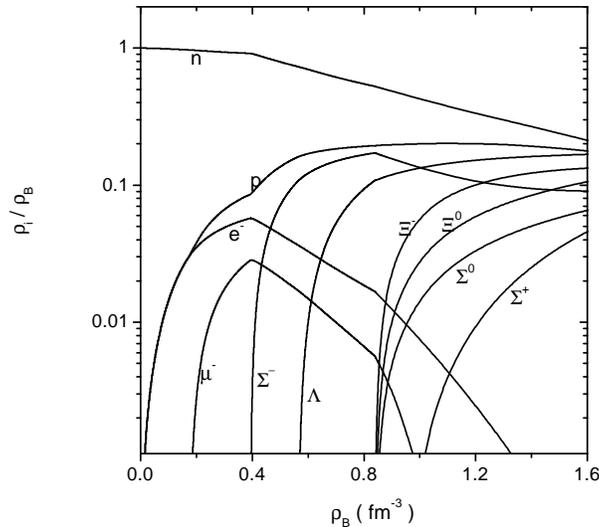}
\caption{The fractions of proton, neutron, $\Lambda$, $\Sigma$ and
$\Xi$ of strange hadronic stars versus baryon density with
$\beta$-equilibrium.}
\end{figure}
\end{center}

\begin{center}
\begin{figure}[hbt]
\includegraphics[scale=0.66]{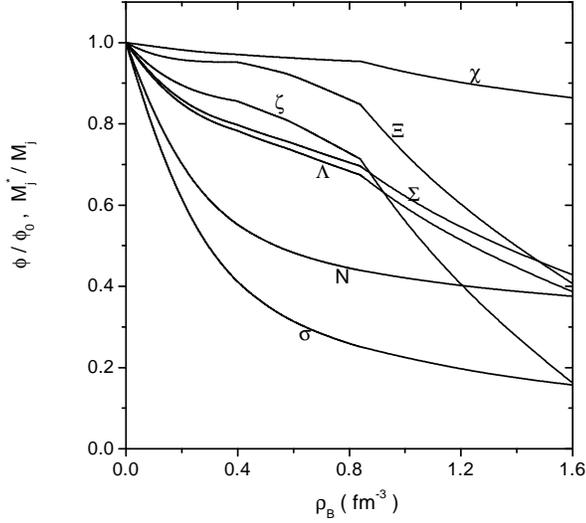}
\caption{The effective baryon masses and meson mean fields versus baryon
density with $\beta$-equilibrium.}
\end{figure}
\end{center}

The density dependence of the effective baryon masses and scalar
mean fields are shown in Fig.~2. The $\sigma$ field decreases quickly
with the increasing baryon density when the density is small,
$\rho_B<0.4$ fm$^{-3}$. This is because at small baryon density,
the nucleon is dominant and there are no hyperons. With the
increasing of density, more and more hyperons appear.
As a result, the $\zeta$ field decreases quickly. At a broad range
of densities, the value of $\chi$ changes little.

In Fig.~3, the pressure versus baryon density is shown. The dashed
and solid lines are for the pure neutron star and the
strange hadronic star with $\beta$-equilibrium, respectively. When
the density is low, the two curves are close to each other.
With the increasing of baryon density, the contributions of protons
and hyperons are not negligible. The inclusion of hyperons will soften
the equation of state of hadronic matter. As a result, at a given baryon
density the pressure of strange hadronic matter is smaller than the
corresponding pressure of pure neutron matter. The pressure $p$ versus
energy density $\varepsilon$ is shown in Fig.~4. Again, one can see that
the equation of state of strange hadronic matter is softer than that of
pure neutron matter.

\begin{center}
\begin{figure}[hbt]
\includegraphics[scale=0.66]{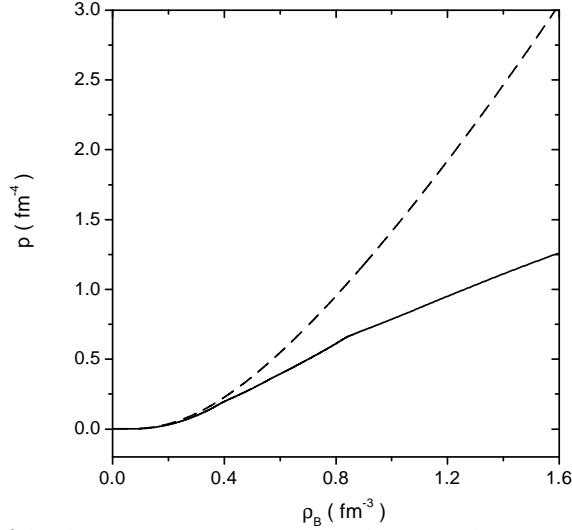}
\caption{The pressure of hadronic matter $p$ versus
baryon density $\rho_B$. The dashed and solid curves are for
pure neutron stars and strange hadronic stars with $\beta$-equilibrium,
respectively.}
\end{figure}
\end{center}

\begin{center}
\begin{figure}[hbt]
\includegraphics[scale=0.66]{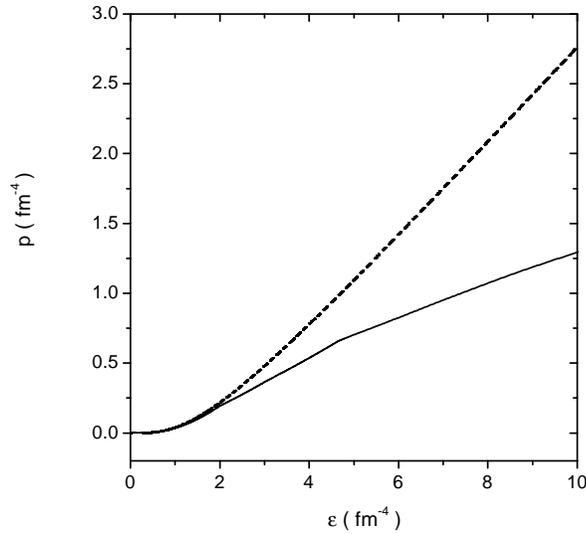}
\caption{The pressure of hadronic matter $p$ versus
energy density $\varepsilon$. The dashed and solid curves are for
pure neutron stars and strange hadronic stars with $\beta$-equilibrium,
respectively.}
\end{figure}
\end{center}

We now study neutron stars with the obtained EOS.
By solving the TOV equation, the baryon density versus radius can be
obtained which is shown in Fig.~5. The central densities $\rho_c$ are chosen
to be $3\rho_0$ and $5\rho_0$ where $\rho_0$ (0.16 fm$^{-3}$) is
the saturation density of symmetric nuclear matter.
The dashed and solid lines are
for pure neutron stars and strange hadronic stars with
$\beta$-equilibrium, respectively. With the increasing radius,
the density of strange hadronic stars decreases a little faster than
that of pure neutron stars which results in a smaller radius.
The radii of stars are not sensitive to their the central
density. For example, for $\rho_c$ of $3\rho_0$ and $5\rho_0$,
the radii are both around 11-12 km.

\begin{center}
\begin{figure}[hbt]
\includegraphics[scale=0.66]{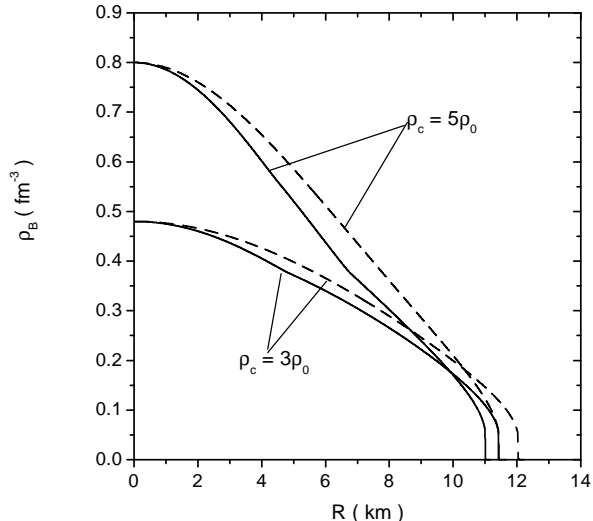}
\caption{The baryon density of a hadronic star versus radius.
The dashed and solid curves are for pure
neutron stars and strange hadronic stars with $\beta$-equilibrium,
respectively.}
\end{figure}
\end{center}

We plot the star mass ratio $M/M_{\mathrm{sun}}$ versus central
baryon density in Fig.~6. The maximum mass of pure neutron stars
is about 1.8$M_{\mathrm{sun}}$ with a central density 1.05
fm$^{-3}$. After the central density is larger than 1.05
fm$^{-3}$, the star will become unstable. The maximum mass changes
to $1.45 M_{\mathrm{sun}}$ when hyperons are included. In the
range $3\rho_0<\rho_c<6\rho_0$, the masses of pure neutron stars
and strange hadronic stars are
$1.48M_{\mathrm{sun}}<M<1.8M_{\mathrm{sun}}$ and
$1.23M_{\mathrm{sun}}<M<1.45M_{\mathrm{sun}}$, respectively. Our results are
reasonable compared with the observation of the six known stars
with masses in the range $1.36 \pm 0.08M_{\mathrm{sun}}$, since the
``neutron star" is in fact a strange hadronic star with
$\beta$-equilibrium. We should also keep in mind that there are
some heavy stars reported in recent years. For PSR J0437-4715, the
mass is found to be $1.58\pm 0.18M_{\mathrm{sun}}$ \cite{Straten}. For Vale
X-1, Cygnus X-2 and 4U 1820-30, their masses are determined to be
$1.87^{+0.23}_{-0.17}M_{\mathrm{sun}}$ \cite{Barziv}, $1.8\pm 0.4 M_{\mathrm{sun}}$
\cite{Orosz} and $\simeq 2.3M_{\mathrm{sun}}$ \cite{Zhang,Miller}. The rotation
of a star can increase its mass by $\sim 10\%$ \cite{rotation}. Therefore, the
calculated maximum mass of strange hadronic stars can be as large as $1.6
M_{\mathrm{sun}}$. If the heavy stars such as 4U 1820-30 are confirmed, the
strange hadronic star would be ruled out if this model is
a good description of Nature.
It is possible to increase the maximum star mass by making the EOS stiffer at
higher densities. Whether the inclusion of a quark core in the
strange star will result in a large-maximum mass is an interesting
topic.

\begin{center}
\begin{figure}[hbt]
\includegraphics[scale=0.66]{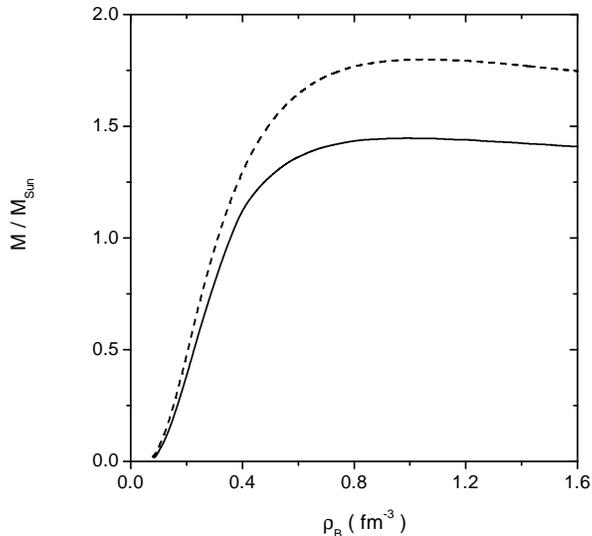}
\caption{The masses of hadronic stars versus their central baryon
densities. The dashed and solid curves are for pure
neutron stars and strange hadronic stars with $\beta$-equilibrium,
respectively.}
\end{figure}
\end{center}

In Fig.~7, the masses of stars versus their radii are shown. For
pure neutron stars, when their masses are in the range
$0.5M_{\mathrm{sun}}<M<1.8M_{\mathrm{sun}}$, their radii are about
11.0-12.3 km. For the strange hadronic stars, when the masses are
in the range $0.5M_{\mathrm{sun}}<M<1.45M_{\mathrm{sun}}$, the radii are about
10.7-11.7 km. With the same mass ($M>0.2M_{\mathrm{sun}}$), strange
hadronic stars have smaller radii compared with pure neutron
stars. Because the size of neutron stars is small, it is very
difficult to observe and measure their radii directly. Different
indirect methods lead to different values of radii with large
errors. For example, for the RX J1856-3754, the radius varies from
5 km to 15 km with a mass of $1.4M_{\mathrm{sun}}$ \cite{Heiselberg}. More
accurate values are needed to obtain a more strict constraint on the
EOS of hadronic matter.

\begin{center}
\begin{figure}[hbt]
\includegraphics[scale=0.66]{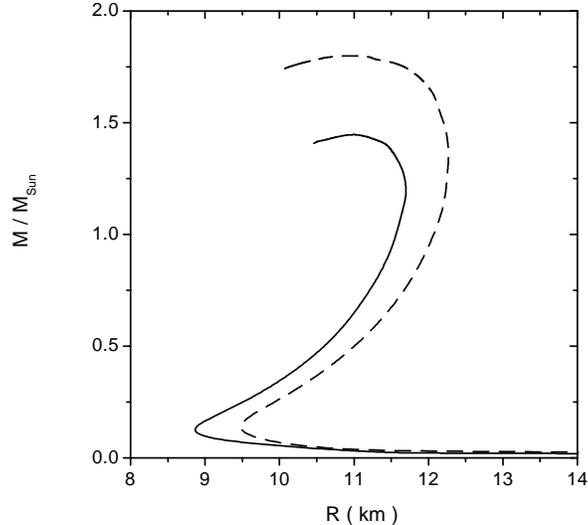}
\caption{The masses of hadronic stars versus their radii.
The dashed and solid curves are for pure
neutron stars and strange hadronic stars with $\beta$-equilibrium,
respectively.}
\end{figure}
\end{center}

\section{summary}

We have investigated pure neutron stars and strange hadronic stars in
the chiral $SU(3)$ quark mean field model. The $\Lambda$, $\Sigma$
and $\Xi$ hyperons are included in the model. The proton and
hyperon contributions to the system are important at high baryon
density when $\beta$-equilibrium is achieved, and soften the
EOS of hadronic matter. The maximum pure neutron star mass is
about $M=1.8M_{\mathrm{sun}}$ with a corresponding radius $R=11.0$ km and
central density $\rho_c=1.05$ fm$^{-3}$. For the strange hadronic
stars, the maximum masses are about $1.45M_{\mathrm{sun}}$ and the
corresponding radii and central density are $R=10.9$ km and
$\rho_c=1.0$ fm$^{-3}$. When the central densities are between
$3\rho_0$ and $6\rho_0$, the masses of stars are in the range
$1.23M_{\mathrm{sun}}<M<1.45M_{\mathrm{sun}}$ (strange hadronic stars) and
$1.48M_{\mathrm{sun}}<M<1.8M_{\mathrm{sun}}$ (pure neutron stars). If the masses of
stars are larger than $0.5M_{\mathrm{sun}}$, the typical values of radii
are 10.7-11.7 km (strange hadronic stars) km and 11.0-12.3 km (pure
neutron stars).

Our results are reasonable compared with astrophysical observations
where the six known neutron stars have masses in the narrow range
$1.36\pm0.08M_{\mathrm{sun}}$. Accurate values of radii for neutron
stars are needed to get a more strict constraint on the EOS of hadronic matter.
As for the heavy stars, for example 4U 1820-30, if its mass
$M\simeq 2.3M_{\mathrm{sun}}$ is confirmed, then strange hadronic
stars are obviously ruled out if the model explored herein is a good
description of Nature.
It is therefore of interest to see whether including quark degrees of
freedom can lead to this large mass.

\section*{Acknowledgements}
P.W. thanks the Theory Group at Jefferson Lab for their kind
hospitality. This work was supported by the Australian Research Council
and by DOE contract DE-AC05-84ER40150, under which SURA operates
Jefferson Laboratory.

\end{document}